\documentclass[cameraready]{Interspeech}

\usepackage{amsmath, amsfonts, amssymb}
\usepackage{graphicx}
\usepackage{booktabs}
\usepackage{siunitx}
\usepackage{subcaption}
\usepackage{enumitem}
\usepackage{multirow}

\title{Whisper-Aware LLM: Self-Supervised Uncertainty Learning for Robust Whispered Speech Recognition}

\author[affiliation={1}]{Gaopeng}{Xu}
\author[affiliation={1}]{Zhenyu}{Wang}
\author[affiliation={1}]{Zheng}{Xue}
\author[affiliation={1}]{Yinfeng}{Xia}
\author[affiliation={1}, correspondingauthor]{Haitao}{Yao}

\address{
$^1$Qwen Business Unit of Alibaba, China
}
\email{\{longque.xgp, wzy497957, jeremy.xz, xiayinfeng.xyf, timmy.yht\}@alibaba-inc.com}


\begin{document}

\maketitle

\begin{abstract}
The signal ambiguity of whispered speech drives ASR systems toward two opposing failure modes: failing to capture whispered speech or hallucinatory transcription of noise. This paper introduces the Whisper-Aware LLM, a framework that teaches an Audio-LLM to perceive and react to this uncertainty. Our model develops an intrinsic self-awareness by learning to quantify the physical deficiencies of acoustic signals through targeted self-supervised tasks. This learned uncertainty is then operationalized via a novel Confidence-Fused Decoding mechanism, which provides both high-level instructions and frame-level attention modulation to the LLM decoder. Our experiments confirm the effectiveness of this approach. The model sets a new state-of-the-art on whispered speech with a 17\% relative CER reduction on AISHELL6-Whisper. At the same time, it directly addresses the reliability trade-off, with hallucination rates dropping from over 25\% to 4.5\%.
\end{abstract}

\section{Introduction}
\label{sec:introduction}

Whispered speech, a mode of phonation without vocal fold vibration, presents a formidable and long-standing challenge for Automatic Speech Recognition (ASR) systems. The acoustic manifestation of whispering characterized by the absence of a fundamental frequency (F0) and harmonic structure, and a noise-like quality \cite{lin2023improving, li2024aishell6} creates a fundamental ambiguity that conventional ASR models struggle to resolve. This ambiguity gives rise to a debilitating accuracy-reliability trade-off. On one hand, models trained primarily on normal speech exhibit perceptual failure when encountering whispers, leading to high error rates. On the other hand, increasing a model's sensitivity to capture these faint cues paradoxically heightens its susceptibility to hallucination, where it erroneously transcribes speech from ambient noise.

Existing research has attempted to mitigate this trade-off via two principal avenues. The first is a data-centric approach focusing on augmentation, such as generating ``pseudo-whisper'' data \cite{lin2023improving, gudepu2020whisper}. While beneficial, these methods are limited by the inevitable distributional gap between synthetic and authentic whispered speech. The second avenue is model-centric static adaptation, exemplified by the projection layer proposed by Li et al. \cite{li2024aishell6}. While an improvement, such ``one-size-fits-all'' static solutions fail to dynamically adjust to the vast spectrum of whisper variability.

Crucially, these prevailing paradigms primarily address the acoustic symptoms of whispered speech, rather than its underlying cause: a fundamental increase in signal uncertainty. A pivotal acoustic study by Lin et al. \cite{lin2023improving} experimentally confirmed that the lack of glottal information is the single most critical factor degrading recognition performance. This key finding suggests a new direction: instead of forcing the model to make high-confidence predictions from highly uncertain signals, a more robust approach is to first teach the model to quantify the signal's intrinsic uncertainty itself.

This paper is built upon this principle. We introduce the Whisper-Aware LLM, a framework designed to endow a native Audio-LLM with an intrinsic self-awareness of signal uncertainty. We posit that a model can learn to perceive this uncertainty in a self-supervised manner by being trained to identify the very physical deficiencies that define whispered speech. To this end, we design two innovative self-supervised tasks: F0 contour prediction and masked spectrum reconstruction. The quantified uncertainty is then operationalized through our Confidence-Fused Decoding mechanism. To ensure stable learning, we introduce a disciplined three-stage training strategy. Our contributions are:
\begin{enumerate}[nosep, leftmargin=*]
    \item A novel framework integrating self-supervised uncertainty perception directly into an Audio-LLM.
    \item Two physics-informed self-supervised tasks for label-free uncertainty learning.
    \item A Confidence-Fused Decoding mechanism for dynamic generative control.
    \item A systematic three-stage training protocol for robust representation learning.
\end{enumerate}

\begin{figure*}[t]
  \centering
  \includegraphics[width=0.9\textwidth]{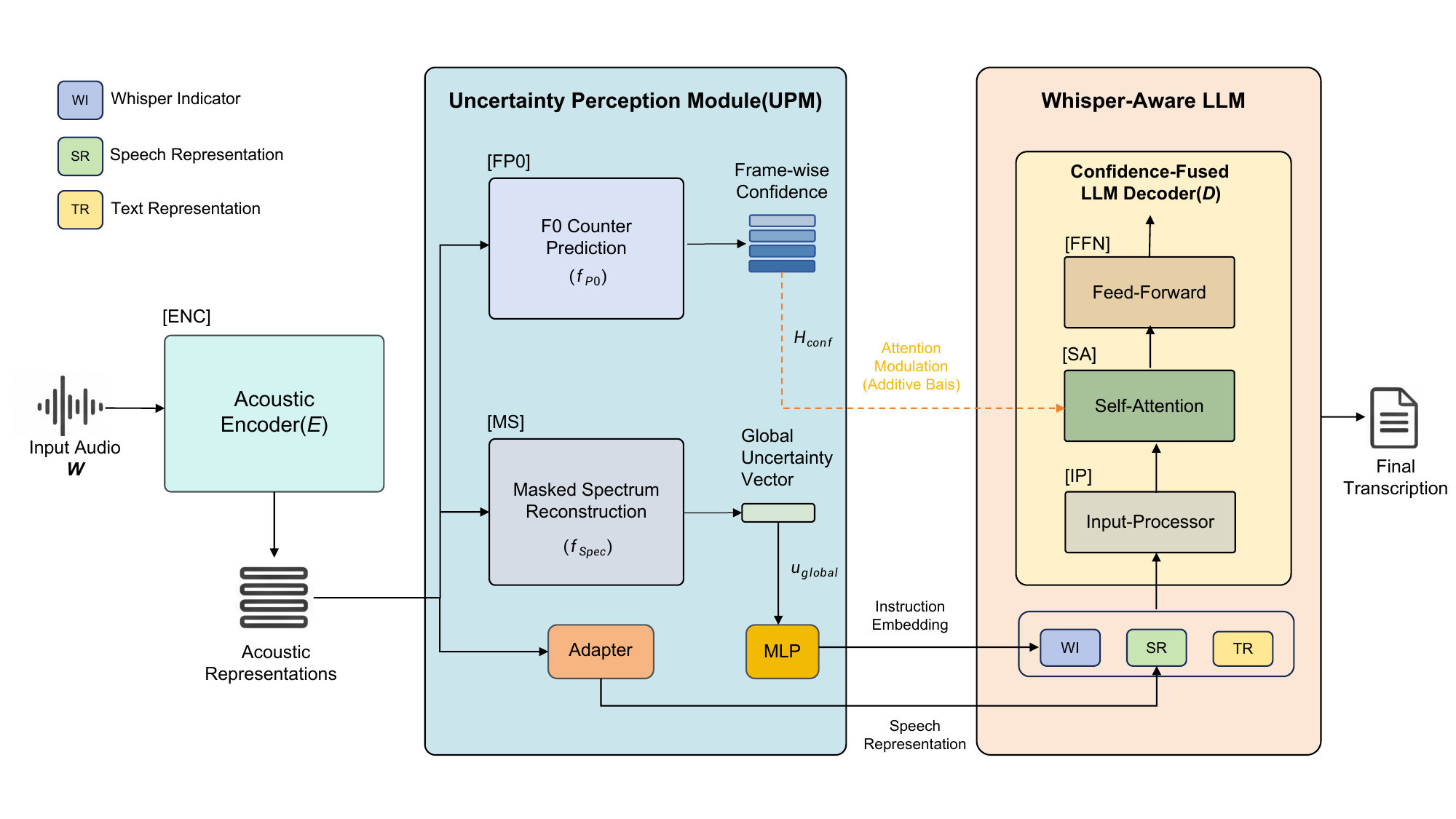}
  \caption{The overall architecture of the Whisper-Aware LLM. The model first encodes the audio into acoustic representations. Our UPM then processes these representations to perceive signal quality, producing a global instruction embedding (\(e_{\text{instruct}}\)) and a frame-wise confidence sequence (\(H_{\text{conf}}\)). These outputs guide the LLM decoder: the instruction acts as a system prompt, and the confidence scores modulate the attention mechanism.}
  \label{fig:system_overview}
\end{figure*}

\section{The Whisper-Aware LLM Framework}
\label{sec:method}

Our framework enhances a standard Audio-LLM by integrating a lightweight Uncertainty Perception Module (UPM) and a corresponding Confidence-Fused Decoding mechanism.

\subsection{Overall Architecture}
As shown in Fig.\ref{fig:system_overview}, the architecture consists of three main stages:
\begin{itemize}
    \item \textbf{Acoustic Encoding:} The encoder $\mathcal{E}$ maps an input audio waveform $w$ into a sequence of representations $H_{\text{enc}} \in \mathbb{R}^{T \times D_{\text{model}}}$.
    \item \textbf{Uncertainty Perception:} Our UPM, $\mathcal{U}$, takes $H_{\text{enc}}$ as input and produces a global uncertainty vector $u_{\text{global}}$ and a frame-wise confidence sequence $H_{\text{conf}}$.
    \item \textbf{Confidence-Fused Decoding:} The decoder $\mathcal{D}$ generates the transcription $Y$ autoregressively, guided by the UPM's outputs.
\end{itemize}

\subsection{Uncertainty Perception Module (UPM)}
The UPM is a small network trained to perceive signal uncertainty through two self-supervised auxiliary tasks.

\subsubsection{Task 1: F0 Contour Prediction}
Motivated by the finding that F0 absence is the most critical factor in whisper recognition degradation \cite{lin2023improving}, we train a prediction head $f_{\text{F0}}$ within the UPM to predict the F0 contour $\hat{F}_0$ from $H_{\text{enc}}$. The training loss is the Mean Squared Error (MSE) against a ground-truth contour $F_0$:
\begin{equation}
    L_{\text{F0}} = \frac{1}{T} \sum_{t=1}^{T} (\hat{F}_0(t) - F_0(t))^2
\end{equation}
The frame-wise confidence score $h_{\text{conf}}(t)$ is derived from the normalized prediction error $e_{\text{F0}}(t)$:
\begin{equation}
    h_{\text{conf}}(t) = 1 - \text{clip}\left(\frac{e_{\text{F0}}(t)}{\epsilon_{\text{max}}}, 0, 1\right)
\end{equation}
where $\epsilon_{\text{max}}$ is an error clipping hyperparameter. The resulting $H_{\text{conf}}$ is then centered to have a zero mean.

\subsubsection{Task 2: Masked Spectrum Reconstruction}
To capture the unstructured, noise-like nature of the whisper spectrum, a second head $f_{\text{Spec}}$ is trained to reconstruct randomly masked frames in $H_{\text{enc}}$. The UPM's final hidden state, averaged over time, forms the global uncertainty vector $u_{\text{global}}$.

\subsection{Confidence-Fused Decoding}

\subsubsection{Global Instruction Embedding}
The global uncertainty vector $u_{\text{global}}$ is transformed by an MLP\cite{rumelhart1986learning}, $f_{\text{instr}}$, into an instruction embedding $e_{\text{instruct}}$. This is prepended to the decoder's input sequence, acting as a high-level system prompt.

\subsubsection{Attention Modulation}
In the LLM-based architecture, the acoustic representations $H_{\text{enc}}$ are treated as a prefix to the token sequence, and the LLM performs causal self-attention over this entire combined sequence. To guide this process, we introduce an attention biasing mechanism. For any query $q$ within the LLM, when it computes an attention score with a key $k_t$ that corresponds to the $t$-th acoustic frame from the encoder, we inject our learned confidence as an additive bias. The modified score is:
\begin{equation}
    \text{score}(q, k_t) = \frac{q^\top k_t}{\sqrt{d_k}} + w \cdot h_{\text{conf}}(t)
    \label{eq:additive_bias}
\end{equation}
where $w$ is a learnable scalar. This simple yet effective mechanism allows the model to dynamically pay less attention to acoustic frames deemed unreliable by the UPM, preventing the decoder from being misled by uncertain evidence.

\subsection{Three-Stage Training Strategy}
To stably integrate our Uncertainty Perception Module (UPM) while avoiding gradient conflicts with the large pre-trained model, we employ a disciplined three-stage training strategy:
\begin{itemize}
    \item \textbf{Stage 1: UPM Pre-training.} We first train the UPM exclusively on its self-supervised objectives ($L_{\text{F0}} + L_{\text{Spec}}$), while the entire Audio-LLM backbone remains frozen. This isolates the UPM, allowing it to rapidly learn robust uncertainty features from the frozen encoder without destabilizing the main model.
    \item \textbf{Stage 2: Interface Adaptation.} Next, we focus on connecting the UPM to the decoder. We jointly train the audio encoder, the audio-LLM adapter, the UPM, and our new decoding interfaces (the MLP $f_{\text{instr}}$ and scalar $w$), while keeping the large LLM decoder frozen. This stage aligns the modalities and teaches the model to use the uncertainty signals.
    \item \textbf{Stage 3: Full End-to-End Fine-tuning.} Finally, the entire model is fine-tuned end-to-end using a composite loss ($L_{\text{ASR}} + \lambda_{\text{aux}}(L_{\text{F0}} + L_{\text{Spec}})$). To maintain efficiency, we apply LoRA\cite{lora} to the LLM decoder.
\end{itemize}

\section{Experiments}
\label{sec:experiments}

To validate the effectiveness of our Whisper-Aware LLM, we conducted a series of comprehensive experiments. We aimed to evaluate its performance on whispered speech recognition, its general ASR capabilities, its robustness against hallucination, and the contribution of its core components.

\begin{table*}[t]
  \caption{Detailed performance comparison on English benchmarks (WER\%/CER\%). Our model demonstrates superior performance on both normal (N) and whispered (W) speech.}
  \label{tab:english_whisper_detailed}
  \centering
  \sisetup{table-format=2.2, round-mode=places, round-precision=2}
  \begin{tabular}{@{\hspace{8pt}}l S[table-format=1.2] S[table-format=1.2] S[table-format=1.2] S[table-format=2.2] | S[table-format=1.2] S[table-format=1.2] S[table-format=1.2] S[table-format=1.2]@{\hspace{8pt}}}
    \toprule
    \multirow{2}{*}{\textbf{Model}} & \multicolumn{4}{c|}{\textbf{WER (\%)}} & \multicolumn{4}{c}{\textbf{CER (\%)}} \\
    \cmidrule(lr){2-5} \cmidrule(l){6-9}
    & {$N_{US}$} & {$N_{SG}$} & {$W_{US}$} & {$W_{SG}$} & {$N_{US}$} & {$N_{SG}$} & {$W_{US}$} & {$W_{SG}$} \\
    \midrule
    S1: Whisper-v3              & 5.25 & 8.67 & 8.20 & 19.03 & 1.03 & 2.50 & 2.47 & 8.85 \\
    S2: Whisper-v3 fine-tuned   & 5.58 & 7.46 & 7.95 & 15.79 & 0.93 & 1.86 & 2.37 & 7.53 \\
    S3: Qwen2-Audio fine-tuned  & 4.95 & 6.85 & 7.40 & 13.50 & 0.88 & 1.65 & 2.30 & 6.20 \\
    S4: Aishell6-whisper AVSR   & 4.50 & 6.06 & 6.35 & 11.63 & 0.73 & 1.33 & 1.95 & 5.17 \\
    S5: Qwen3-ASR\cite{qwen3}   & 4.99 & 7.26 & 7.01 & 16.46 & 0.98 & 1.85 & 2.45 & 7.11 \\
    S6: Funasr-ASR\cite{fun}    & 4.65 & 6.96 & 8.48 & 19.29 & 0.91 & 1.76 & 2.88 & 8.24 \\
    S7: Seed-ASR\cite{seed}     & 4.29 & 6.51 & 6.88 & 12.95 & 0.85 & 1.62 & 2.15 & 6.03 \\
    \midrule
    \textbf{Ours}               & \bfseries 4.15 & \bfseries 5.89 & \bfseries 5.92 & \bfseries 10.81 & \bfseries 0.70 & \bfseries 1.25 & \bfseries 1.85 & \bfseries 4.98 \\
    \bottomrule
  \end{tabular}
\end{table*}

\subsection{Experimental Setup}

\subsubsection{Datasets and Metrics}

\textbf{Training Datasets.} Our work leverages two strategically constructed datasets for different training phases to build a highly robust model.
\begin{itemize}
    \item \textit{UPM Pre-training Corpus (for Stage 1):} To ensure the UPM learns a comprehensive model of acoustic signals, we constructed a large-scale, mixed-domain corpus. This corpus includes General-Purpose Speech (a 3,000h subset of WenetSpeech\cite{wenetspeech}, a 3,000h subset of GigaSpeech\cite{gigaspeech}, AISHELL-1\cite{aishell1} and LibriSpeech\cite{librispeech}), Whispered Speech (wTIMIT\cite{wtimit}, AISHELL6-Whisper\cite{aishell6}), and a 1,000h collection of pure noise.
    \item \textit{Fine-tuning Set (for Stage 2 \& 3):} This set consists of the training splits of AISHELL-1, LibriSpeech, wTIMIT, AISHELL6-Whisper, and a 200h subset of the pure noise data to maintain robustness.
\end{itemize}

\textbf{Evaluation Datasets.} We evaluate our model's performance across three key dimensions:
\begin{itemize}
    \item \textit{Whispered Speech:} The official test sets of AISHELL6-Whisper \cite{li2024aishell6} (Chinese) and wTIMIT \cite{lim2011computational} (English).
    \item \textit{General-Purpose ASR:} The test sets of AISHELL-1 and LibriSpeech to assess general competency.
    \item \textit{Reliability and Hallucination:} To specifically evaluate hallucination under adverse conditions, we created a Noise Hallucination Set containing 1000 audio clips. These clips consist of various non-speech sounds (e.g., machine hum, wind) and segments with either no speech or very faint speech heavily masked by noise, designed to be challenging for conventional VADs and prone to trigger false transcriptions.
\end{itemize}

\textbf{Metrics.} The primary performance metrics are Character Error Rate (CER\%) for Chinese and Word Error Rate (WER\%)/CER\% for English. On the Noise Hallucination Set, we measure the Hallucination Rate (HR\%), calculated as the percentage of audio files containing no transcribable speech that produce a non-empty transcription.

\subsubsection{Implementation Details}
\begin{itemize}
    \item \textbf{Base Model:} Our framework is built upon the Qwen2-Audio model \cite{qwen2a}. It consists of an audio encoder, a lightweight projector as the audio-LLM adapter, and a Qwen-7B LLM \cite{qwen} backbone.
    \item \textbf{Uncertainty Perception Module (UPM):} Our UPM is composed of a shared feature extractor (two 1D-CNN layers with GeLU and a single Transformer\cite{ts} encoder layer) followed by two task-specific heads for F0 prediction and masked spectrum reconstruction.
    \item \textbf{Training Details:} Our three-stage strategy was implemented using the AdamW\cite{ad} optimizer with a learning rate of \SI{1e-5}{}. In Stage 3, the LLM was updated via LoRA (rank=64, alpha=32). The auxiliary loss weight was set to $\lambda_{\text{aux}}=0.1$. Stage 1 ran for 100k steps, while Stages 2 and 3 each ran for 20k steps.
\end{itemize}

\subsubsection{Compared Systems}
We benchmark against internal baselines sharing the Qwen2-Audio backbone and external state-of-the-art (SOTA) systems. The systems S1-S7 are defined as: Whisper-v3 \cite{whisper} (S1) and its fine-tuned version (S2); a fine-tuned Qwen2-Audio baseline (S3); and other recent SOTA ASR systems (S4-S7).

\subsection{Main Results}

\subsubsection{Whispered Speech Accuracy}
We first evaluate whispered speech accuracy on both Chinese and English benchmarks to demonstrate the cross-lingual effectiveness of our framework.

As detailed in Table \ref{tab:chinese_whisper}, our model achieves a new state-of-the-art on the challenging AISHELL6-Whisper test set. With a CER of only 1.31\% on whispered speech, it significantly surpasses all compared systems, including the strong Seed-ASR (S7) which scored 1.58\%. This represents a relative CER reduction of 17\% over the previous best. Importantly, our model also achieves the lowest error rate on normal speech (0.63\%), indicating that its specialization for whispers does not compromise its general ASR capabilities.

\begin{table}[h]
  \caption{Performance on the AISHELL6-Whisper set (CER\%).}
  \label{tab:chinese_whisper}
  \centering
  \sisetup{table-format=2.2, round-mode=places, round-precision=2}
  \begin{tabular}{@{\hspace{6pt}}l S S@{\hspace{6pt}}}
    \toprule
    \multirow{2}{*}{\textbf{Model}} & \multicolumn{2}{c}{\textbf{CER (\%)}} \\
    \cmidrule(l){2-3}
    & {Normal} & {Whisper} \\
    \midrule
    S1: Whisper-v3                 & 3.95 & 18.93 \\
    S2: Whisper-v3 fine-tuned      & 1.62 & 6.69  \\
    S3: Qwen2-Audio fine-tuned     & 1.05 & 3.98  \\
    S4: Aishell6-whisper AVSR      & 1.11 & 4.13  \\
    S5: Qwen3-ASR                  & 0.64 & 3.79  \\
    S6: Funasr-ASR                 & 0.72 & 19.50 \\
    S7: Seed-ASR                   & 0.65 & 1.58  \\
    \midrule
    \textbf{Ours}                  & \bfseries 0.63 & \bfseries 1.31 \\
    \bottomrule
  \end{tabular}
\end{table}

The results on English benchmarks, presented in Table \ref{tab:english_whisper_detailed}, further reinforce these findings. Our model consistently outperforms all baselines across both normal (N) and whispered (W) conditions for US and Singaporean (SG) accents. Notably, while other models show a dramatic increase in WER when transitioning from normal to whispered speech, our model exhibits a much more graceful degradation. This demonstrates its enhanced robustness to the acoustic shift caused by whispering, a direct result of its uncertainty-aware architecture.

\subsubsection{General ASR Performance}
To explicitly verify that our model's specialization for whispers does not impair its general capabilities, we summarize its performance on standard, non-whispered benchmarks in Table \ref{tab:general_asr}. Our model demonstrates highly competitive performance against state-of-the-art systems. On the AISHELL-1 test set, it achieves a strong CER of 1.34\%, placing it on par with other top-tier systems like Funasr (1.22\%). On the LibriSpeech-clean set, our model obtains a robust WER of 1.91\%, again showing performance in the same league as leading models. This confirms that our uncertainty-aware framework successfully enhances robustness in challenging conditions without sacrificing its proficiency on high-quality, normal speech.

\begin{table}[h]
  \caption{General ASR performance on standard benchmarks.}
  \label{tab:general_asr}
  \centering
  \sisetup{table-format=1.2, round-mode=places, round-precision=2}
  \begin{tabular}{@{\hspace{10pt}}l S S@{\hspace{10pt}}}
    \toprule
    \textbf{Model} & \multicolumn{1}{c}{\textbf{AISHELL-1}} & \multicolumn{1}{c}{\textbf{LibriSpeech-clean}} \\
                   & \multicolumn{1}{c}{(CER\%)}            & \multicolumn{1}{c}{(WER\%)}                    \\
    \midrule
    Whisper-v3 & 4.72 & 1.86 \\
    Seed-ASR   & 1.63 & 2.80 \\
    Funasr     & 1.22 & 1.51 \\
    \midrule
    \textbf{Ours} & \bfseries 1.34 & \bfseries 1.91 \\
    \bottomrule
  \end{tabular}
\end{table}

\subsubsection{Reliability and Hallucination}
A core contribution of our work is resolving the accuracy-reliability trade-off. We test this on our curated Noise Hallucination Set, where an ideal system should produce no output. As shown in Table \ref{tab:reliability}, standard fine-tuned models exhibit a high Hallucination Rate (HR). For instance, even strong baselines like S7 and S5 incorrectly generate text for over 25\% of the files, with the highest rate reaching 29.3\% (S6). This highlights their tendency to ``imagine'' speech when faced with ambiguous signals. In stark contrast, our Whisper-Aware LLM achieves an HR of only 4.5\%, a substantial reduction. This remarkable improvement demonstrates that by learning to perceive signal uncertainty, our model effectively learns when not to speak, drastically mitigating the risk of generating spurious text from non-speech audio. This reduction is primarily attributed to the Global Instruction mechanism, which guides the decoder, as quantified in our ablation study.

\begin{table}[h]
  \caption{Reliability evaluation on the Noise Hallucination Set.}
  \label{tab:reliability}
  \centering
  \sisetup{table-format=2.1, round-mode=places, round-precision=1}
  \begin{tabular}{@{\hspace{6pt}}l S@{\hspace{6pt}}}
    \toprule
    \textbf{Model}  & {HR(\%)} \\
    \midrule
    S3: Qwen2-Audio fine-tuned & 35.5  \\
    S5: Qwen3-ASR              & 25.2  \\
    S6: Funasr-ASR             & 29.3  \\
    S7: Seed-ASR               & 25.2 \\
    \midrule
    \textbf{Ours}              & \bfseries 4.5 \\
    \bottomrule
  \end{tabular}
\end{table}

\subsection{Ablation of Model Components}
To dissect the contribution of our proposed components, we conducted an ablation study on the AISHELL6-Whisper, with results in Table \ref{tab:ablation_components}. Starting from the S3 baseline (3.98\% CER), using only the frame-wise Attention Modulation yields a clear improvement to 3.45\%. The Global Instruction, however, proves to be the primary driver of performance, slashing the CER to 1.84\% by providing high-level, strategic awareness of signal quality. Finally, the Full Model combines both, achieving the best performance at 1.31\%.

\begin{table}[th]
  \caption{Ablation of Confidence-Fused Decoding components on AISHELL6-Whisper (Whisper subset).}
  \label{tab:ablation_components}
  \centering
  \sisetup{table-format=1.2, round-mode=places, round-precision=2}
  \begin{tabular}{@{\hspace{6pt}}l S@{\hspace{6pt}}}
    \toprule
    \textbf{Configuration} & {CER(\%)}  \\
    \midrule
    Baseline (Vanilla Finetuning) & 3.98  \\
    + Attention Modulation Only & 3.45  \\
    + Global Instruction Only & 1.84  \\
    \textbf{+ Full Model (Ours)} & \bfseries 1.31  \\
    \bottomrule
  \end{tabular}
\end{table}

\section{Conclusion}
\label{sec:conclusion}

In this paper, we introduced the Whisper-Aware LLM, a framework designed to handle the uncertainty inherent in whispered speech. Instead of just matching patterns, our approach teaches the model to first perceive signal quality through targeted self-supervised tasks. This learned awareness is then used by our Confidence-Fused Decoding mechanism to guide the LLM, allowing it to selectively trust the acoustic evidence it receives. Our experiments confirm the success of this strategy. The model achieves state-of-the-art results on whispered speech benchmarks and drastically reduces hallucinations, all while remaining a highly effective general-purpose ASR system.

\section{Generative AI Use Disclosure}
The authors used generative AI tools solely for language editing and grammar polishing. All technical content, methodology, experimental design, and analysis were conducted entirely by the authors.

\bibliographystyle{IEEEtran}
\bibliography{mybib}

\end{document}